\documentclass[12pt]{iopart}

\usepackage{iopams}  

\expandafter\let\csname equation*\endcsname\relax
\expandafter\let\csname endequation*\endcsname\relax

\usepackage{epsfig, amsmath, amssymb}
\newtheorem{theorem}{Theorem}%  meant for continuous numbers
\newtheorem{prop}{Proposition}
\newtheorem{lemma}{Lemma}
\newtheorem{corollary}[prop]{Corollary}

\newtheorem{assumption}{Assumption}
\usepackage{hyperref}
\usepackage{lineno}

\usepackage{xcolor}
\usepackage{booktabs}
\usepackage{multirow}
\usepackage{cite}

\begin{document}

\title[Continual Quantum Architecture Search with Tensor-Train Encoding]{Continual Quantum Architecture Search with Tensor-Train Encoding: Theory and Applications to Signal Processing}

\author{Jun Qi$^{1*}$, Chao-Han Huck Yang$^{2}$, Pin-Yu Chen$^{3}$, Javier Tejedor$^{4}$, Ling Li$^{5*}$, Min-Hsiu Hsieh$^{6*}$}

\address{1. School of Electrical and Computer Engineering, Georgia Institute of Technology, Atlanta, GA 30332, USA     \\ 
2. NVIDIA Research, Santa Clara, CA 95051, USA	 \\
3. IBM Thomas J. Watson Research Center, NY, 10598, USA    \\
4. Department of Information Technology, Institute of Technology, Universidad San Pablo-CEU, CEU universities, Urbanización Montepríncipe, Boadilla del Monte, 28668, Spain \\
5. City St George's, University of London, England, EC1V 0HB, United Kingdom \\
6. Hon Hai (Foxconn) Quantum Computing Research Center, Taipei, 114, Taiwan   \\
}
\ead{jqi41@gatech.edu, caroline.li@city.ac.uk, minhsiuh@gmail.com}
\hspace{25mm}\small{* denotes corresponding authors}
\vspace{10pt}

%\linenumbers

\begin{abstract}
We introduce CL-QAS, a continual quantum architecture search framework that mitigates the challenges of costly amplitude encoding and catastrophic forgetting in variational quantum circuits. The method uses Tensor-Train encoding to efficiently compress high-dimensional stochastic signals into low-rank quantum feature representations. A bi-loop learning strategy separates circuit parameter optimization from architecture exploration, while an Elastic Weight Consolidation regularization ensures stability across sequential tasks. We derive theoretical upper bounds on approximation, generalization, and robustness under quantum noise, demonstrating that CL-QAS achieves controllable expressivity, sample-efficient generalization, and smooth convergence without barren plateaus. Empirical evaluations on electrocardiogram (ECG)-based signal classification and financial time-series forecasting confirm substantial improvements in accuracy, balanced accuracy, F1 score, and reward. CL-QAS maintains strong forward and backward transfer and exhibits bounded degradation under depolarizing and readout noise, highlighting its potential for adaptive, noise-resilient quantum learning on near-term devices.
\end{abstract}

%\noindent{\it Keywords}: X, Y, Z
%
% Uncomment for Submitted to journal-title message
%\maketitle

\section{Introduction}
\label{sec1}

Quantum machine learning (QML)~\cite{cerezo2022challenges, power_data, biamonte2017quantum} has emerged as a promising paradigm that leverages quantum circuits to enhance representation power and computational efficiency in data-driven tasks. Among the most widely studied models, variational quantum circuits (VQCs)~\cite{cerezo2021variational, kandala2017hardware, benedetti2019parameterized, qi2025tensorhyper} have shown strong potential for classification and signal processing problems on Noisy Intermediate-Scale Quantum (NISQ) devices~\cite{preskill2018quantum, bharti2022noisy}. However, two persistent challenges limit their scalability and practical deployment. First, amplitude encoding, a common approach for mapping classical data into quantum states, requires exponentially many coefficients and quantum gates, making it computationally expensive for high-dimensional inputs~\cite{schuld2021effect, huang2021power}. Second, when tasks arrive sequentially, VQCs trained in a conventional setting suffer from catastrophic forgetting~\cite{zenke2017continual}, where performance on earlier tasks deteriorates after learning new ones. Addressing these bottlenecks is essential for advancing quantum learning in real-world scenarios such as electrocardiogram (ECG) signal analysis~\cite{rnmo2006electrocardiogram, sanei2013eeg} and financial time-series prediction~\cite{yan2018financial}. 

In this work, we propose CL-QAS, a continual learning framework~\cite{hadsell2020embracing} that integrates structural compression, architectural flexibility, and stability mechanisms for adaptive quantum learning. At its core, CL-QAS employs tensor-train (TT)-based~\cite{oseledets2011tensor} amplitude encoding to factorize high-dimensional inputs into compact tensor networks, thereby reducing state-preparation cost while maintaining expressive capacity. To overcome the rigidity of fixed circuit structures, we introduce a bi-loop quantum architecture search (QAS)~\cite{zhang2022differentiable, du2022quantum} scheme, in which the inner loop optimizes VQC parameters for each task. In contrast, the outer loop adaptively explores circuit architectures while accounting for hardware constraints. To mitigate forgetting, we incorporate elastic weight consolidation (EWC)~\cite{huszar2018note} into the training process, which regularizes parameter updates and preserves information learned from previous tasks. 

\begin{figure}[t]
\centerline{\epsfig{figure=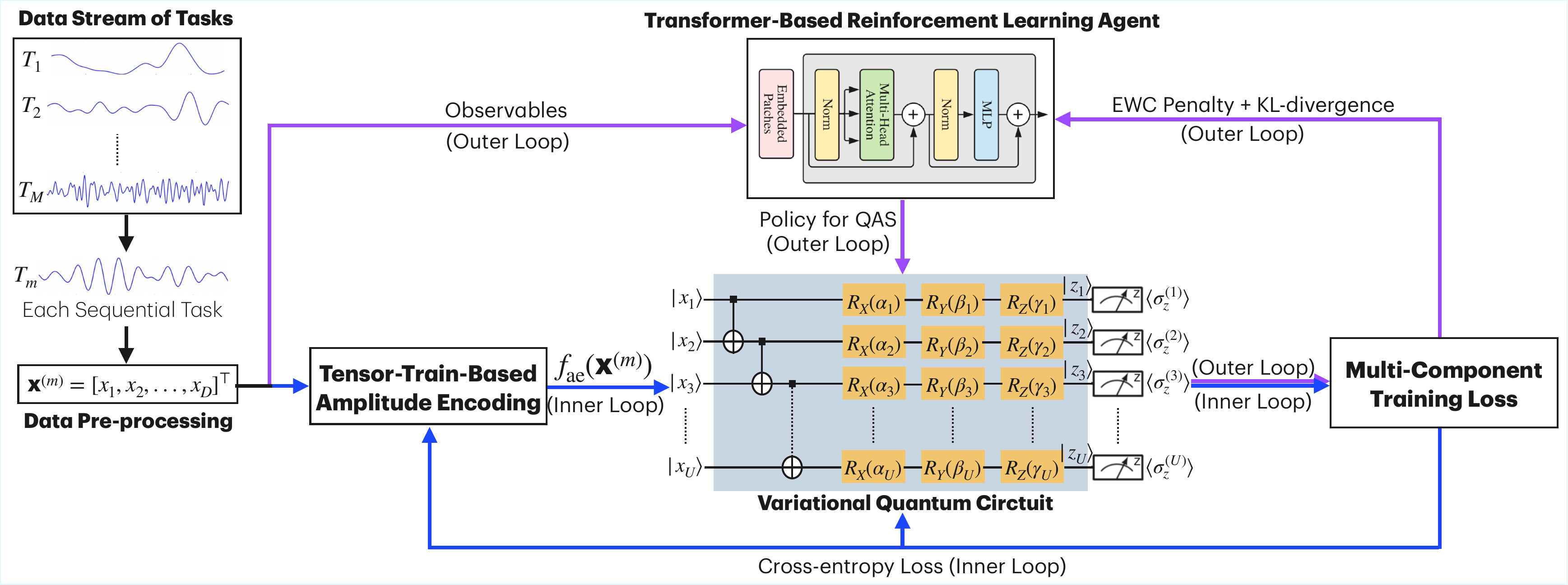, width=155mm}}
\caption{{\it Fundamental architecture of the CL-QAS framework}. The framework consists of three integrated modules: (i) a continual learning engine that incorporates EWC and hardware-aware penalties in the outer loop, (ii) a Transformer-based reinforcement learning policy that generates adaptive gate sequences for VQCs, and (iii) a VQC classifier trained via cross-entropy loss in the inner loop. Sequential tasks (such as ECG or financial signals) are pre-processed and encoded into quantum states using TT amplitude encoding. During training, the QNN parameters are optimized with respect to a fixed circuit architecture. At the same time, the policy network is updated using detached rewards, enabling adaptive search over quantum architectures across tasks.}
\label{fig:cl_qas}
\end{figure}

Beyond algorithmic design, we establish theoretical guarantees on the expressivity, generalization capability, and trainability~\cite{bach2024learning, mohri2018foundations} of CL-QAS. Our analysis reveals that TT encoding regulates representational capacity through low-rank factorization, QAS constrains circuit depth while enhancing adaptability, and EWC introduces curvature-aware regularization that accelerates convergence and stabilizes continual learning. Collectively, these mechanisms ensure polynomial scaling in sample complexity and gradient variance, in contrast to the exponential dependence often observed in conventional amplitude-encoded VQCs. 

We empirically validate CL-QAS across two representative signal-processing applications (ECG-based signal classification and financial time-series forecasting) to demonstrate its efficiency and generalization. On ECG datasets, CL-QAS improves classification accuracy and mitigates catastrophic forgetting across sequential tasks compared with baseline VQCs. On financial data, the framework achieves superior balanced accuracy and F1 score while requiring shallower circuits and fewer two-qubit gates, confirming its potential for resource-efficient quantum learning on both simulated and real IBM Heron hardware. 

In summary, this work makes three primary contributions:
\begin{itemize}
\item We introduce CL-QAS, a continual-learning quantum framework that unifies TT-based amplitude encoding, bi-loop quantum architecture search, and EWC regularization.
\item We derive analytical bounds on the representation capacity, trainability, and generalization performance of CL-QAS, establishing improved scalability and stability over conventional VQCs. 
\item We validate CL-QAS across simulated and hardware experiments on ECG and financial datasets, demonstrating consistent improvements in accuracy, generalization, and hardware efficiency. 
\end{itemize}

\section{Results}
\label{res}

\subsection{CL-QAS Methodology}
\label{sec:method}

The CL-QAS framework integrates reinforcement-learning-driven circuit search with continual-learning regularization to design adaptive, hardware-feasible VQCs. As shown in Figure~\ref{fig:cl_qas}, each task $m$ proceeds in two stages: an inner-loop fixed-architecture training stage and an outer-loop policy-reinforcement stage. In the inner loop, a VQC generated by a policy network is optimized through cross-entropy loss~\cite{mao2023cross} on dataset $\mathcal{D}_m$. In the outer loop, the architecture-generated policy is updated using a REINFORCE-style gradient estimator stabilized by EWC and Kullback–Leibler (KL) divergence regularization. In particular, the entire CL-QAS pipeline comprises three main components: a TT-based amplitude-encoding module, a trainable VQC classifier, and a transformer-based reinforcement learning policy equipped with continual learning regularizers.

The TT-based amplitude encoding compresses the classical input vectors into compact quantum states. Given the input dimension $D$ and qubit count $U$, for an input feature vector $\textbf{x} \in \mathbb{R}^{D}$, the amplitude vector is shown in Eq. (\ref{eq:tt}). 
\begin{equation}
\label{eq:tt}
\textbf{x}[i_1, i_2, ..., i_{D}] = \sum\limits_{r_1, ..., r_{U}} \mathcal{G}_{i_1, r_1}^{(1)} \mathcal{G}_{r_1, i_2, r_2}^{(2)} \cdot\cdot\cdot \mathcal{G}_{i_{U-1}, r_U}^{(U)},
\end{equation}
where each TT-core $\mathcal{G}^{(u)}$ has a maximum TT-rank $r$. This reduces the memory cost from $\mathcal{O}(2^{U})$ to $\mathcal{O}(Ur^2)$ and yields a normalized quantum state $\vert \textbf{x} \rangle = \vert x_1 \rangle \otimes \vert x_2 \rangle \otimes \cdot\cdot\cdot \otimes \vert x_{U} \rangle$, thereby making amplitude encoding scalable without exponential overhead. 

Encoded states are then processed by a parameterized VQC composed of single-qubit rotations and entangling CNOT gates, as presented in Eq. (\ref{eq:ent}).
\begin{equation}
\label{eq:ent}
\mathcal{U}(\boldsymbol{\theta}) = \prod\limits_{l=1}^{L} \left( \prod\limits_{u=1}^{U} R_{X}(\alpha_{u}^{(l)}) R_{Y}(\beta_{u}^{(l)}) R_{Z}(\gamma_{u}^{(l)}) \prod\limits_{u=1}^{U-1} \text{CNOT}(u,u+1) \right),
\end{equation}
where $L$ denotes the depths of the VQC, and $\boldsymbol{\theta}$ denotes the variational parameters. The operators $R_{X}(\cdot)$, $R_{Y}(\cdot)$, and $R_{Z}(\cdot)$ are trainable rotation gates, while $\text{CNOT}$$(u, u+1)$ introduces entanglement. Measurement in the Pauli-Z basis produces an expectation vector $\boldsymbol{\sigma} = [\langle \sigma_{z}^{(1)} \rangle, \langle \sigma_{z}^{(2)} \rangle, ..., \langle \sigma_{z}^{(U)} \rangle]^{\top}$, which is transformed into class probabilities via a softmax layer, as shown in Eq. (\ref{eq:softmax}).
\begin{equation}
\label{eq:softmax}
\hat{\textbf{y}} = \text{softmax}\left( \boldsymbol{\sigma}[1:K]	\right), \hspace{3mm} (K < U), 
\end{equation}
where $K$ denotes the number of classes in the task. 

The reinforcement learning module governs the adaptive search for circuit architectures. The policy network proposes a candidate architecture $\mathcal{A}_m$ with probability $\pi_{\boldsymbol{\phi}}(\mathcal{A}_m)$, using validation accuracy $c$ as a reward and penalizing excessive entanglement. The REINFORCE update is presented in Eq. (\ref{eq:j1}).
\begin{equation}
\label{eq:j1}
\nabla_{\boldsymbol{\phi}}J(\boldsymbol{\phi}) = -\mathbb{E}_{\mathcal{A}_{m} \sim \pi_{\boldsymbol{\phi}}} \left[ c \nabla_{\boldsymbol{\phi}}\log \pi_{\boldsymbol{\phi}}(\mathcal{A}_{m}) \right],
\end{equation}
with the policy objective presented in Eq. (\ref{eq:j2}).
\begin{equation}
\label{eq:j2}
J(\boldsymbol{\phi}) = -\mathbb{E}_{\mathcal{A}_{m} \sim \pi_{\boldsymbol{\phi}}} \left[ c \log\pi_{\boldsymbol{\phi}}(\mathcal{A}_{m}) \right]. 
\end{equation}

To ensure knowledge retention and policy stability, CL-QAS applies two regularization mechanisms. The first is the EWC penalty, which constrains important parameters to remain near their previously learned values via Eq. (\ref{eq:omega}).
\begin{equation}
\label{eq:omega}
\Omega_{\rm ewc}(\boldsymbol{\phi}, \boldsymbol{\phi}^{\rm (old)}) = \frac{\lambda}{2} \sum\limits_{i} \mathcal{F}_{i}(\boldsymbol{\phi}_{i} - \boldsymbol{\phi}_{i}^{\rm (old)})^{2},
\end{equation}
where $\mathcal{F}_i$ represents Fisher-information entry for parameter $\phi_i$, and $\lambda$ controls regularization strength. The second mechanism is a KL-divergence term promoting similarity between the current policy distribution $\pi_{\boldsymbol{\phi}}$ and a prior $\pi_{\rm prior}$, as shown in Eq. (\ref{eq:kl}).
\begin{equation}
\label{eq:kl}
\Omega_{\rm kl}(\pi_{\boldsymbol{\phi}}, \pi_{\rm prior}) = \text{KL}(\pi_{\boldsymbol{\phi}} \vert\vert \pi_{\rm prior}). 
\end{equation}

The optimization of CL-QAS follows a bi-loop scheme that separates circuit-parameter learning from policy adaptation. In the inner loop, for a fixed architecture $\mathcal{A}_m$, the VQC parameters $\boldsymbol{\theta}$ are optimized via cross-entropy loss, as presented in Eq. (\ref{eq:vqc}).
\begin{equation}
\label{eq:vqc}
\mathcal{L}_{\rm vqc}(\boldsymbol{\theta}; \mathcal{A}_{m}) = -\frac{1}{\vert \mathcal{D}_{m} \vert} \sum\limits_{(\textbf{x}, y)\in \mathcal{D}_m} \log p_{\boldsymbol{\theta}}(y \vert \textbf{x}, \mathcal{A}_m), 
\end{equation}
while in the outer loop, policy parameters $\boldsymbol{\phi}$ are updated using the detached reward signal, regularized by continual learning and KL-divergence terms, as denoted in Eq. (\ref{eq:policy}).
\begin{equation}
\label{eq:policy}
\mathcal{L}_{\rm policy}(\boldsymbol{\phi}) = J(\boldsymbol{\phi}) + \mu \Omega_{\rm ewc}(\boldsymbol{\phi}, \boldsymbol{\phi}^{\rm (old)}) + \beta \Omega_{\rm kl}(\pi_{\boldsymbol{\phi}}, \pi_{\rm prior}),
\end{equation}
where $\mu$ and $\beta$ weight the two contributions. The overall joint objective combining the two loops is presented in Eq. (\ref{eq:cl_qas}).
\begin{equation}
\label{eq:cl_qas}
\mathcal{L}_{\rm \text{cl-qas}}(\boldsymbol{\theta}, \boldsymbol{\phi}) = \sum\limits_{m=1}^{M}\mathcal{L}_{\rm vqc}(\boldsymbol{\theta}; \mathcal{A}_{m}) + \lambda \mathcal{L}_{\rm policy}(\boldsymbol{\phi}), 
\end{equation}
with the number of tasks $M$ and $\lambda$ balancing circuit training and policy adaptation. This formulation enables CL-QAS to learn adaptive, hardware-feasible quantum circuits while continually preserving prior task knowledge.

\subsection{Theoretical Results of CL-QAS}
\label{sec:thm}

We analyze the theoretical properties of CL-QAS from four perspectives: (1) TT-based amplitude-encoding expressivity, (2) bi-loop trainability, (3) overall generalization, and (4) robustness against quantum noise. To establish the results, we begin with the following assumption and lemma that formalize the smoothness and stability conditions of the learning framework. 

\begin{assumption}[Smoothness and Bounded Gradients]. 
\label{assp}
The per-task VQC loss function $\mathcal{L}_{\rm vqc}(\boldsymbol{\theta}; \mathcal{A}_{m})$ is twice continuously differentiable and $L_{\boldsymbol{\theta}}$-smooth, Eq. (\ref{eq:lip}) stands. 
\begin{equation}
\label{eq:lip}
\left\lVert \nabla_{\boldsymbol{\theta}}\mathcal{L}_{\rm vqc}(\boldsymbol{\theta}_{1}; \mathcal{A}_m) - \nabla_{\boldsymbol{\theta}}\mathcal{L}_{\rm vqc}(\boldsymbol{\theta}_{2}; \mathcal{A}_m) \right\rVert_{2} \le L_{\boldsymbol{\theta}} \left\lVert \boldsymbol{\theta}_1 - \boldsymbol{\theta}_2 \right\rVert_{2}. 
\end{equation}
For a bounded curvature constant $C_{\rm curv}$, the policy objective $J(\boldsymbol{\phi})$ further satisfies $\Vert \nabla_{\boldsymbol{\phi}}^{2} J(\boldsymbol{\phi}) \rVert_{2} \le C_{\rm curv}$, and stochastic gradient obeys $\mathbb{E}\lVert \nabla_{\boldsymbol{\theta}} \mathcal{L}_{\rm vqc}(\boldsymbol{\theta}) \rVert_{2}^{2} \le G^{2}$ for some constant $G$. 
\end{assumption}

\begin{lemma}[Continual Learning Stability]. 
\label{lem}
If the policy objective function $J(\boldsymbol{\phi})$ is $C_{\rm curv}$-smooth, then for any update from $\boldsymbol{\phi}^{(old)}$ to $\boldsymbol{\phi}$, Eq. (\ref{eq:curv}) stands. 
\begin{equation}
\label{eq:curv}
\left\vert J(\boldsymbol{\phi}) - J(\boldsymbol{\phi}^{\rm (old)}) \right\vert \le C_{\rm curv}\left\lVert \boldsymbol{\phi} - \boldsymbol{\phi}^{\rm (old)} \right\rVert_{2}^{2},
\end{equation}
quantifying the stability of EWC-regularized policy updates. 
\end{lemma}

\textit{TT-based Amplitude-Encoding Expressivity}. We next analyze the fidelity of TT-based amplitude encoding. When classical data vectors $\textbf{x} \in \mathbb{R}^{2^{U}}$ are mapped into normalized quantum states, storing these states scales exponentially with the input dimension. TT decomposition provides a compact representation that approximates $\textbf{x}$ with 
complexity governed by the TT-rank $r$. Let $\tilde{\textbf{x}}$ denote the TT approximation obtained by TT-based Singular Value Decomposition (TT-SVD)~\cite{lee2015estimating}, with the approximation error as shown in Eq. (\ref{eq:err_tt}).
\begin{equation}
\label{eq:err_tt}
\epsilon_{\rm tt} = \lVert \textbf{x} - \tilde{\textbf{x}} \rVert_{2} \le \left(	\sum\limits_{u=1}^{U-1} \sum\limits_{j > r} s_{u, j}^{2} \right)^{\frac{1}{2}}, 
\end{equation}
where $s_{u, j}$ presents the singular values discarded during decomposition. After normalization of $\textbf{x}$ and $\hat{\textbf{x}}$, we obtain the normalized terms $\hat{\textbf{x}}$ and $\hat{\tilde{\textbf{x}}}$ as separately shown in Eq. (\ref{eq:norm_x}).
\begin{equation}
\label{eq:norm_x}
\hat{\textbf{x}} = \frac{\textbf{x}}{\lVert \textbf{x} \rVert_{2}}, \hspace{2mm} \text{and} \hspace{2mm} \hat{\tilde{\textbf{x}}} = \frac{\tilde{\textbf{x}}}{\lVert \tilde{\textbf{x}} \rVert_{2}}.
\end{equation}

The fidelity between their corresponding quantum states is bounded in Theorem~\ref{thm:fidelity}. 

\begin{theorem}[Fidelity Bound].
\label{thm:fidelity}
If $\rho = \frac{\epsilon_{\rm tt}}{\lVert \textbf{x} \rVert_{2}} < 1$, then we have Eq. (\ref{eq:fidelity}).
\begin{equation}
\label{eq:fidelity}
\mathcal{F}(\hat{\textbf{x}}, \hat{\tilde{\textbf{x}}}) = \vert \langle \hat{\textbf{x}}, \hat{\tilde{\textbf{x}}} \rangle \vert^{2} \ge \left( \frac{1 - \rho}{1 + \rho}	\right)^{2} \ge 1 - \mathcal{O}\left(\frac{1}{r}\right).
\end{equation}
\end{theorem}

Hence, increasing the TT-rank $r$ improves fidelity at the expense of higher computational cost, thereby governing the fidelity-compression trade-off in CL-QAS.

\textit{Bi-Loop Trainability of CL-QAS}. We then examine trainability. The success of VQC classifiers relies on maintaining non-vanishing gradients and stable updates across sequential tasks. CL-QAS addresses this by employing a bi-loop optimization structure that couples circuit and policy updates. 

\begin{theorem}[Trainability Guarantee of CL-QAS]. 
\label{thm:train}
Under smooth losses and bounded Fisher curvature $C_{\rm curv}$, the expected squared gradient variance norm satisfies the upper bound in Eq. (\ref{eq:thm1}).
\begin{equation}
\label{eq:thm1}
\mathbb{E}\left[	\left\lVert \nabla_{\boldsymbol{\theta}, \boldsymbol{\phi}} \mathcal{L}_{\rm \text{cl-qas}}(\boldsymbol{\theta}, \boldsymbol{\phi}) \right\rVert_{2}^{2} \right] \le \frac{C_1 \sigma^{2}}{\vert D_{m} \vert} + \frac{C_2}{r^{2}} + C_{3} C_{\rm curv}\left\lVert \boldsymbol{\phi} - \boldsymbol{\phi}^{(old)} \right\rVert_{2}^{2}, 
\end{equation}
where $\sigma^{2}$ refers to the variance of $\mathcal{L}_{\rm vqc}$, and $r$ is the maximum TT-rank. $C_1$, $C_2$, and $C_3$ separately depend on smoothness and gradient-variance from Assumption~\ref{assp}, TT-induced Neural Tangent Kernel (NTK)~\cite{bietti2019inductive, liu2022representation, jacot2018neural} deviation, and the EWC weight $\lambda$.
\end{theorem}

In particular, in Theorem~\ref{thm:train}, the bound shows that CL-QAS maintains polynomially bounded gradients, $\mathcal{O}(1/\vert D_{m} \vert) + \mathcal{O}(1/r^{2})$, mitigating exponential vanishing typical of dense encoders. The bound yields the expressivity-trainability trade-off summarized below. 

\begin{corollary}[Expressivity-Trainability Trade-off]. 
Given data-dependent constants $C_{i} > 0$, or any target tolerance $\eta \in (0, 1)$, Eq. (\ref{eq:choice}) is represented as follows. 
\begin{equation}
\label{eq:choice}
r \ge \frac{C_2}{\eta}, \hspace{4mm} \vert D_{m} \vert \ge \frac{C_1}{\eta^{2}},
\end{equation} 
which ensure simultaneous fidelity and trainability satisfying $\mathbb{E}\lVert \nabla_{\boldsymbol{\theta}, \boldsymbol{\phi}} \mathcal{L}_{\rm cl-qas}(\boldsymbol{\theta}, \boldsymbol{\phi}) \rVert_{2}^{2} \le \eta$. 
\end{corollary}

Moreover, minimizing the proxy bound $\frac{C_1\sigma^2}{\vert D_{m} \vert} + \frac{C_2}{r^2}$ under fixed compute cost yields the maximum TT-rank as shown in Eq. (\ref{eq:rr}).
\begin{equation}
\label{eq:rr}
r^{*} \asymp \sqrt{\vert D_m \vert},
\end{equation}
implying that larger datasets support proportionally smaller TT-ranks without loss of convergence.

\textit{The Overall Generalization of CL-QAS}. We further formalize the generalization behavior by analyzing the inner-loop VQC and outer-loop policy simultaneously. 

\begin{lemma}[Inner-Loop Generalization]. 
\label{lem:inner}
For bounded parameter norm $\lVert \boldsymbol{\theta} \rVert_{2} \le B$ and the dataset $D_m$, with probability at least $1 - \delta$, Eq. (\ref{eq:lvqc}) stands. 
\begin{equation}
\label{eq:lvqc}
\mathcal{L}_{\rm vqc}(\boldsymbol{\theta}; \mathcal{A}_m) \le \hat{\mathcal{L}}_{\rm vqc}(\boldsymbol{\theta}; \mathcal{A}_m) + \frac{2 B}{\sqrt{\vert D_m \vert}}  +  \sqrt{\frac{\log(1/\delta)}{2 \vert D_m \vert}},
\end{equation}
where $\hat{\mathcal{L}}_{\rm vqc}(\cdot)$ is the empirical risk, yielding standard Hoeffding–Bernstein bounds for bounded models. 
\end{lemma}

\begin{lemma}[Outer-Loop Probably Approximately Correct (PAC)-Bayes Bound]. 
\label{lem:outer}
For policy distribution $\pi_{\boldsymbol{\phi}}$ and prior $\pi_{\rm prior}$, for any $\delta \in (0, 1)$, Eq. (\ref{eq:j3}) stands. 
\begin{equation}
\label{eq:j3}
J(\boldsymbol{\phi}) \le \hat{J}(\boldsymbol{\phi}) + \sqrt{\frac{\text{KL}(\pi_{\boldsymbol{\phi}} \vert\vert \pi_{\rm prior}) + \log(\frac{1}{\delta})}{2 \vert D_{m} \vert}}, 
\end{equation}
with the empirical policy objective function $\hat{J}(\cdot)$ associated with the expected one $J(\cdot)$. The upper bound controls the architecture search policy's generalization gap. 
\end{lemma}

\begin{theorem}[Overall Generalization Bound]. 
\label{thm:overall}
Combining Assumption~\ref{assp} (smoothness and bounded gradients), Lemma~\ref{lem:inner} (Inner-Loop Generalization), and Lemma~\ref{lem:outer} (Outer-Loop PAC-Bayes Bound), then, for a constant $C_{\rm tt} > 0$, with probability at least $1 - \delta$ over the dataset $\mathcal{D}_m$ and policy trajectories drawn from $\pi_{\boldsymbol{\phi}}$, Eq. (\ref{eq:thm4}) stands. 
\begin{equation}
\label{eq:thm4}
\begin{split}
\mathcal{L}_{\rm cl-qas}(\boldsymbol{\theta}, \boldsymbol{\phi}) &\le \hat{\mathcal{L}}_{\rm vqc}(\boldsymbol{\theta}; \mathcal{A}_m) + \frac{2 B}{\sqrt{\vert D_m \vert}}  + \sqrt{\frac{\text{KL}(\pi_{\boldsymbol{\phi}} \vert\vert \pi_{\rm prior}) + \log(1/ {\delta})}{2 \vert D_{m} \vert}} \\
&\hspace{4mm} + C_{\rm tt}\epsilon_{\rm tt}  + C_{\rm curv}\lVert	 \boldsymbol{\phi} - \boldsymbol{\phi}^{\rm (old)} \rVert_{2}. 
\end{split}
\end{equation}
\end{theorem}

This bound formally characterizes the interplay among expressivity (through TT fidelity), stability (through EWC curvature control), and generalization (through PAC-Bayes regularization), providing theoretical guarantees for scalable and stable learning in CL-QAS. 

\textit{Robustness Against Quantum Noise}. We finally analyze the robustness of CL-QAS to quantum noise. For task $m$, $z_{m}(\textbf{x}; \boldsymbol{\theta}, \mathcal{A}_m) \in [-1, 1]^{U}$ denotes the noiseless vector of Pauli-Z expectations, and $\textbf{z}_{m}^{\mathcal{N}}$ presents its value after a noisy channel $\mathcal{N}$. Thus, the noise model is composed of (1) single-qubit depolarizing with probability $p_{1}$; (2) two-qubit depolarizing on entanglers with probability $p_2$; (3) symmetric readout flips with probability $p_r$; and (4) encoder-angle jitter $\epsilon$ added to the TT-encoded angles. 

Let $n_{1,m}$ and $n_{2,m}$ be the counts of one- and two-qubit gates used by $\mathcal{A}_m$ in the forward pass. We assume the standard contraction modes as shown in Eq. (\ref{eq:model1}). 
\begin{equation}
\label{eq:model1}
\zeta_1, \zeta_2 \in [0, 1], \hspace{2mm} \zeta_1 = 1 - \frac{4}{3}p_1, \hspace{2mm} \zeta_2 = 1 - \frac{16}{15}p_2,
\end{equation}

Swapping to $\zeta_i = 1 - p_i$ results in Eq. (\ref{eq:zeta}).
\begin{equation}
\label{eq:zeta}
\alpha_m = \zeta_{1}^{n_{1,m}} \zeta_{2}^{n_{2,m}} (1 - 2p_{r}), \hspace{4mm} \textbf{z}_{m}^{\mathcal{N}} = \alpha_{m} \textbf{z}_{m} + \boldsymbol{\Delta}_{m}, 
\end{equation}
with a remainder $\boldsymbol{\Delta}_{m}$ that aggregates mixing terms and encoder jitter, which further yields Eq. (\ref{eq:jitter}). 
\begin{equation}
\label{eq:jitter}
\lVert \boldsymbol{\Delta}_m \rVert_{2} \le \delta_{m}. 
\end{equation} 

Then, the per-example classifier loss $\ell(\cdot, y)$ is $L_{\ell}$-Lipschitz and $G$-smooth, so that, we obtain the objective-level robustness in Theorem~\ref{thm:robust} by combining per-task classifier robustness in Lemma~\ref{lem:pertask} and policy-term robustness in Lemma~\ref{lem:policy}. 

\begin{lemma}[Per-task classifier robustness].
\label{lem:pertask}
For each task $m$, Eq. (\ref{eq:per_task}) stands.
\begin{equation}
\label{eq:per_task}
\left\vert \mathcal{L}_{\rm vqc}^{\mathcal{N}}(\boldsymbol{\theta}; \mathcal{A}_{m}) - \mathcal{L}_{\rm vqc}(\boldsymbol{\theta}; \mathcal{A}_m) \right\vert \le L_{\ell} \left[ (1-\alpha_{m}) \mathbb{E}\lVert \textbf{z}_{m} \rVert_{2} + \delta_{m} \right] \le L_{\ell}\sqrt{U} \left[ (1-\alpha_{m}) + \delta_{m} \right], 
\end{equation}
where $\mathcal{L}_{\rm vqc}^{\mathcal{N}}(\cdot)$ denotes the loss function in the noisy channel $\mathcal{N}$. 
\end{lemma}

\begin{lemma}[Policy-term robustness]. 
\label{lem:policy}
Let the policy objective $J(\boldsymbol{\phi})$ be computed from a validation reward $c^{\mathcal{N}}, c \in [0, 1]$. Given a threshold constant $\epsilon_{c} > 0$ and $\bar{\alpha} \in [0, 1]$, Eq. (\ref{eq:cn}) stands.
\begin{equation}
\label{eq:cn}
\vert c^{\mathcal{N}} - \bar{\alpha} c \vert \le \epsilon_{c}. 
\end{equation}
Then, we further derive Eq. (\ref{eq:lln}) as follows:
\begin{equation}
\label{eq:lln}
\left\vert \mathcal{L}^{\mathcal{N}}_{\rm policy}(\boldsymbol{\phi}) - \mathcal{L}_{\rm policy}(\boldsymbol{\phi}) 	\right\vert \le C_{\pi}\left[ (1 - \bar{\alpha}) + \epsilon_{c} \right], 
\end{equation}
where $\mathcal{L}^{\mathcal{N}}_{\rm policy}$ denotes the policy loss function over the noisy channel $\mathcal{N}$, and $C_{\pi}$ depends on the reward scale and policy score-function moments. 
\end{lemma}

\begin{theorem}[Objective-level robustness].
\label{thm:robust}
By combining Lemma~\ref{lem:pertask} and Lemma~\ref{lem:policy}, with $C_{m} = L_{\ell} \sqrt{U}$, Eq. (\ref{eq:rob}) stands. 
\begin{equation}
\label{eq:rob}
\left\vert \mathcal{L}_{\rm cl-qas}^{\mathcal{N}}(\boldsymbol{\theta}, \boldsymbol{\phi}) - \mathcal{L}_{\rm cl-qas}(\boldsymbol{\theta}, \boldsymbol{\phi}) \right\vert \le \sum\limits_{m=1}^{M} C_{m}\left[ (1 - \alpha_{m}) + \delta_{m} \right] + \lambda C_{\pi}[ (1 - \bar{\alpha}) + \epsilon_{c}],
\end{equation}
where we consider a total of $M$ tasks, and $\mathcal{L}_{\rm cl-qas}^{\mathcal{N}}$ denotes the loss function of CL-QAS over the noisy channel $\mathcal{N}$. 
\end{theorem}

\subsection{Experimental Results}
\label{sec:exp}

To validate the effectiveness of the proposed CL-QAS framework, we conduct experiments on two representative classes of non-stationary signal-processing tasks: ECG-based brain–computer interface (BCI) classification~\cite{huan2004neural} and financial time-series prediction~\cite{van2001financial}. These domains exhibit evolving data distributions, sequential task structures, and high-noise regimes—precisely the conditions under which continual quantum learning is most beneficial. 

In the noiseless and simulated-noise conditions, all experiments compare three core models: a fixed-architecture variational quantum circuit (Naive-VQC), a reinforcement-learning-based quantum architecture search without continual regularization (QAS-No-CL), and the proposed CL-QAS model, which integrates reinforcement-driven architecture adaptation with continual-learning regularization through EWC. In contrast, for the real-hardware experiments, CL-QAS is compared against two advanced baselines: TTN+VQC~\cite{qi2023theoretical}, which combines a TT network with a variational quantum circuit, and Differential QAS~\cite{zhang2022differentiable}, which employs differential gradients to guide the architecture search process. Particularly, Differential QAS learns continuous architecture weights jointly with circuit parameters, which are then discretized for final evaluation. Unlike CL-QAS, it does not use EWC as a regularizer for continual learning. 

Each model employs amplitude encoding on 20 qubits for ECG classification and financial prediction, with a $6$-layer variational circuit composed of single-qubit rotations and entangling CNOT gates. Training uses cross-entropy loss and the Adam optimizer (learning rate of $3\times 10^{-3}$, batch size of $128$, and $100$ epochs). The number of quantum measurements is set to $1024$ for all VQC models. The reinforcement learning reward combines classification accuracy with a mild penalty proportional to the number of CNOT gates, encouraging hardware-efficient circuits. All metrics are averaged across eight sequential tasks and five random seeds, and reported as mean $\pm$ standard deviation. 

\textit{ECG Signal Classification}. The first experiment evaluates CL-QAS on an ECG-based classification dataset derived from the MIT-BIH Arrhythmia Database~\cite{apandi2018arrhythmia}. It comprises eight sequential tasks (Normal vs. Ventricular beats) emulating inter-subject and session-wise variability. A TT encoder is employed with input modes $(4,16,4)$, output $(3,2,2)$, and ranks $(1,2,3,1)$, compressing each $256$-dimensional input into a compact $20$-qubit amplitude state while maintaining temporal-spatial structure and noise-resilient feature encoding. 

To make the comparison across tasks fair, each record is split into training/validation/test subsets in a stratified, time-agnostic manner, with an 80/10/10 split. Concretely, we first split the beats into $80\%$ for training and $20\%$ for held-out evaluation, while preserving the N/V class proportions; the $80\%$ training split is further divided into $85\%$ training and $15\%$ validation. This yields, per task, roughly $68\%$ data for actual parameter updates, $12\%$ for model selection/reward computation in QAS, and $20\%$ for reporting test metrics. The same split is used for all methods so that performance differences can be attributed to the learning algorithm rather than to data leakage or resampling bias. All results in Table~\ref{tab:res1}--\ref{tab:res3} are averaged over the eight tasks and five random seeds to account for record-level variability and the stochasticity of the policy search. 

Under noiseless simulation, as shown in Table~\ref{tab:res1}, CL-QAS consistently outperforms both baselines across all metrics. The average F1-score of $0.819$ exceeds that of Naive-VQC $(0.741 \pm 0.052)$ and QAS-No-CL $(0.748\pm 0.045)$, corresponding to relative gains of $10.5\%$ and $9.5\%$, respectively. Although accuracies are similar across methods, CL-QAS yields higher F1 Scores and rewards, indicating a better precision-recall balance and reduced forgetting. 

Under simulated quantum noise in Table~\ref{tab:res2} ($0.1\%$ depolarizing, $0.1\%$ two-qubit Pauli, $1\%$ readout error), CL-QAS maintains the highest resilience (Acc$=0.932\pm 0.015$, bAcc$=0.936\pm 0.012$, F1$=0.729\pm 0.041$) compared with Naive-VQC (F1$=0.713 \pm 0.048$) and QAS-No-CL (F1$=0.688 \pm 0.058$). The combination of TT-encoding and EWC regularization constrains the hypothesis space and smooths gradient dynamics, prevents over-entanglement, and enables robust optimization even in noisy regimes. 

To further validate practical deployment, hardware experiments were conducted on IBM's $156$-qubit Heron r$2$ processor (Table~\ref{tab:res3}). CL-QAS demonstrates the best mean performance (Acc$=0.863\pm0.008$, bAcc$=0.900\pm0.060$, F1$=0.496\pm0.009$, Rwd$=0.767\pm0.002$), outperforming both TTN+VQC (Acc$=0.846\pm0.011$, bAcc$=0.881\pm0.070$, F1$=0.481\pm0.012$) and Differential QAS (Acc$=0.822\pm0.012$, bAcc$=0.878\pm 0.080$, F1$=0.437\pm0.010$). These hardware-level results confirm that continual regularization and TT compression jointly mitigate decoherence, readout noise, and stochastic gate fluctuations, achieving noise-robust learning on real NISQ systems. 

\begin{table}[t]
\centering
\scriptsize
\caption{Performance comparison of Naive-VQC, QAS-No-CL, and CL-QAS systems on ECG signal classification across 8 tasks given 5 seeds. Our experimental simulations are conducted without quantum noise. The results are reported by means of accuracy (Acc), balanced accuracy (bAcc), F1-score (F1), and Rewards (Rwd).}
\label{tab:res1}
\renewcommand{\arraystretch}{1.1}
\setlength{\tabcolsep}{4pt}
\begin{tabular}{lcccc|cccc|cccc}
\toprule
\multirow{2}{*}{Task} & 
\multicolumn{4}{c|}{\textbf{Naive-VQC}} & 
\multicolumn{4}{c|}{\textbf{QAS-No-CL}} & 
\multicolumn{4}{c}{\textbf{CL-QAS}} \\
\cmidrule(lr){2-5}\cmidrule(lr){6-9}\cmidrule(l){10-13}
& Acc & bAcc & F1 & Rwd & Acc & bAcc & F1 & Rwd & Acc & bAcc & F1 & Rwd \\
\midrule
1 & 0.975 & 0.987 & 0.714 & --- & 0.981 & 0.990 & 0.769 & 0.4264 & 0.981 & 0.990 & 0.769 & 0.9397 \\
2 & 0.944 & 0.969 & 0.769 & --- & 0.876 & 0.932 & 0.600 & 0.9326 & 0.944 & 0.969 & 0.769 & 0.8497 \\
3 & 0.888 & 0.780 & 0.182 & --- & 0.870 & 0.770 & 0.160 & 0.6719 & 0.988 & 0.830 & 0.667 & 0.5347 \\
4 & 1.000 & 1.000 & 1.000 & --- & 1.000 & 1.000 & 1.000 & 0.9104 & 1.000 & 1.000 & 1.000 & 1.0000 \\
5 & 0.938 & 0.967 & 0.615 & --- & 0.938 & 0.967 & 0.615 & 0.9104 & 0.956 & 0.977 & 0.696 & 0.9604 \\
6 & 1.000 & 1.000 & 1.000 & --- & 1.000 & 1.000 & 1.000 & 0.9363 & 1.000 & 1.000 & 1.000 & 0.9500 \\
7 & 0.851 & 0.876 & 0.774 & --- & 0.901 & 0.917 & 0.840 & 0.8624 & 0.851 & 0.876 & 0.774 & 0.9225 \\
8 & 0.988 & 0.993 & 0.875 & --- & 1.000 & 1.000 & 1.000 & 1.0000 & 0.988 & 0.993 & 0.875 & 0.9268 \\
\midrule
Mean & 0.948 & 0.947 & 0.741 & --- & 0.946 & 0.947 & 0.748 & 0.8313 &  0.963 & 0.955 & 0.819 &  0.8855 \\
\midrule 
Deviation & 0.014 & 0.013 & 0.052 & --- & 0.018 & 0.015 & 0.045 & 0.0777 & 0.009 & 0.011 & 0.031 & 0.0384 \\
\bottomrule
\end{tabular}
\end{table}

\begin{table}[t]
\centering
\scriptsize
\caption{Performance comparison of Naive-VQC, QAS-No-CL, and CL-QAS systems on ECG signal classification under depolarizing and readout noise across $8$ tasks given $5$ seeds. Reported results include accuracy (Acc), balanced accuracy (bAcc), F1-score (F1), and Rewards (Rwd). Noise model includes depolarizing noise ($0.1\%$), two-qubit Pauli errors ($0.1\%$), and readout error ($1\%$).}
\label{tab:res2}
\renewcommand{\arraystretch}{1.1}
\setlength{\tabcolsep}{4pt}
\begin{tabular}{lcccc|cccc|cccc}
\toprule
\multirow{2}{*}{Task} & 
\multicolumn{4}{c|}{\textbf{Naive-VQC}} & 
\multicolumn{4}{c|}{\textbf{QAS-No-CL}} & 
\multicolumn{4}{c}{\textbf{CL-QAS}} \\
\cmidrule(lr){2-5}\cmidrule(lr){6-9}\cmidrule(l){10-13}
& Acc & bAcc & F1 & Rwd & Acc & bAcc & F1 & Rwd & Acc & bAcc & F1 & Rwd \\
\midrule
1 & 0.981 & 0.990 & 0.769 & --- & 0.385 & 0.683 & 0.092 & 0.9397 & 0.975 & 0.987 & 0.714 & 0.8936 \\
2 & 0.925 & 0.959 & 0.714 & --- & 0.944 & 0.969 & 0.769 & 0.9532 & 0.901 & 0.945 & 0.652 & 0.9532 \\
3 & 0.702 & 0.685 & 0.077 & --- & 0.938 & 0.805 & 0.286 & 0.2750 & 0.975 & 0.824 & 0.500 & 0.2500 \\
4 & 0.994 & 0.997 & 0.947 & --- & 0.994 & 0.997 & 0.947 & 0.2500 & 1.000 & 1.000 & 1.000 & 1.0000 \\
5 & 0.938 & 0.967 & 0.615 & --- & 0.938 & 0.967 & 0.615 & 0.3000 & 0.938 & 0.967 & 0.615 & 0.8768 \\
6 & 1.000 & 1.000 & 1.000 & --- & 1.000 & 1.000 & 1.000 & 0.3000 & 0.838 & 0.898 & 0.711 & 1.0000 \\
7 & 0.876 & 0.893 & 0.804 & --- & 0.870 & 0.889 & 0.796 & 0.8338 & 0.845 & 0.872 & 0.766 & 0.9242 \\
8 & 0.975 & 0.987 & 0.778 & --- & 1.000 & 1.000 & 1.000 & 0.3000 & 0.988 & 0.993 & 0.875 & 1.0000 \\
\midrule
Mean   & 0.924 & 0.935 & 0.713 & ---  &0.883 & 0.914 & 0.688 & 0.5190 & 0.932 & 0.936 & 0.729 & 0.8622 \\
\midrule
Deviation & 0.018 & 0.014 & 0.048 & --- &0.026 & 0.021 & 0.058 & 0.0342 & 0.015 & 0.012 & 0.041 & 0.0259 \\
\bottomrule
\end{tabular}
\end{table}

\begin{table}[t]
\centering
\scriptsize
\caption{Experimental results on a 156-qubit IBM quantum processor (Heron r2) comparing CL-QAS with TTN+VQC and Differential QAS on ECG signal classification across $8$ tasks given $5$ seeds. Reported results include accuracy (Acc), balanced accuracy (bAcc), F1-score (F1), and Rewards (Rwd).}
\label{tab:results}
\begin{tabular}{lcccc}
\toprule
\textbf{Method} & \textbf{Acc} & \textbf{bAcc} & \textbf{F1} & \textbf{Reward} \\
\midrule
TTN+VQC & $0.846 \pm 0.011$ & $0.881 \pm 0.070$ & $0.481 \pm 0.012$ & N/A \\
\midrule
Differential QAS & $0.822 \pm 0.012$ & $0.878 \pm 0.080$ & $0.437 \pm 0.010$ & $0.5409 \pm 0.0021$ \\
\midrule
CL-QAS & $\mathbf{0.863 \pm 0.008}$ & $\mathbf{0.900 \pm 0.060}$ & $\mathbf{0.496 \pm 0.009}$ & $\mathbf{0.7674 \pm 0.0018}$ \\
\bottomrule
\end{tabular}
\label{tab:res3}
\end{table}

Finally, to assess the contribution of tensor-train (TT) encoding, we further conducted an ablation study comparing CL-QAS (with TT encoding) and CL-QAS (without TT encoding) on the signal classification tasks. As summarized in Table~\ref{tab:ablation_tt}, TT encoding consistently improves classification performance with less runtime overhead. The TT structure enhances feature expressivity and noise robustness, resulting in higher balanced accuracy and reward than the non-TT variant.

\begin{table}[h]
\centering
\scriptsize
\caption{Ablation study comparing CL-QAS with and without TT-encoding on ECG signal classification. Results are reported as mean $\pm$ standard deviation over eight tasks and five seeds. Reported results include accuracy (Acc), balanced accuracy (bAcc), F1-score (F1), Rewards (Rwd), and Runtime. Noise model includes depolarizing noise ($0.1\%$), two-qubit Pauli errors ($0.1\%$), and readout error ($1\%$).}
\label{tab:ablation_tt}
\begin{tabular}{lccccc}
\toprule
\textbf{Method} & \textbf{Acc} & \textbf{bAcc} & \textbf{F1} & \textbf{Reward (Rwd)} & \textbf{Runtime} \\
\midrule
CL-QAS (TT)    & $0.932 \pm 0.015$ & $0.936 \pm 0.012$ & $0.729 \pm 0.041$ & $0.8622 \pm 0.2359$ & $38.0 \pm 6.0$s \\
CL-QAS (No-TT) & $0.922 \pm 0.009$ & $0.924 \pm 0.014$ & $0.714 \pm 0.053$ & $0.8478 \pm 0.1989$ & $43.5 \pm 6.8$s \\
\bottomrule
\end{tabular}
\end{table}

\textit{Financial Time-Series Forecasting}. We further assess CL-QAS on a synthetic–realistic financial time-series benchmark that exhibits regime shifts and temporal dependencies. Starting from one or more close-price series (or a synthetic AR-driven price with changing drift/volatility), we compute eight technical-indicator channels per time step—returns, rolling mean/volatility, RSI, MACD components, momentum, and Bollinger-band z-scores—and stack the most recent $32$ steps, producing a $256$-dimensional feature vector ($32 \times 8 = 256$) for every decision point. Accordingly, the TT encoder factorizes the 256-dimensional input as $(4,16,4)\rightarrow (5,2,2)$ with TT-ranks $(1,2,3,1)$, producing $20$ rotation angles per circuit. The label of each step is the next-step market direction (up vs. down). The entire time axis is then partitioned into eight consecutive, non-overlapping segments, each corresponding to a distinct market regime and thereby forming a single CL task.

Since the data are time-ordered, we use a chronological split within each regime: the earliest $80\%$ of samples are used to train the policy-selected VQC, the next $10\%$ for validation/reward estimation, and the last $10\%$ for testing. No future information is leaked into the past. This mirrors a realistic deployment in which a QML model must adapt to new regimes without re-accessing historical market states. The same $80/10/10$ time split is applied to all baselines (Naive-VQC and QAS-No-CL) and to CL-QAS. Reported metrics in Table~\ref{tab:finance_1}--\ref{tab:finance_ibm}, including accuracy, balanced accuracy, F1-score, and reward, are averaged over the eight regimes and five seeds to smooth out regime-specific volatility and to make the IBM-hardware comparisons statistically meaningful. 

In noiseless conditions (Table~\ref{tab:finance_1}), CL-QAS achieves the best overall performance (Acc$=0.604\pm 0.063$, bAcc$=0.604\pm 0.070$, F1$=0.624\pm 0.063$, Rwd$= 0.6054\pm 0.0622$), surpassing Naive-VQC and QAS-No-CL. When realistic quantum noise is introduced in Table~\ref{tab:finance_2}, all methods degrade modestly but retain relative ordering; CL-QAS (Acc$=0.637\pm 0.110$, bAcc$=0.641\pm 0.096$, F1$=0.618\pm 0.168$, Rwd$=0.6395\pm 0.0882$) maintains high stability due to its noise-aware policy regularization. 

Hardware deployment on the IBM Heron R2 device, as shown in Table~\ref{tab:finance_ibm}, further demonstrates the framework's resilience: CL-QAS attains the best performance (Acc$=0.607\pm0.015$, bAcc$=0.607\pm0.011$, F1$=0.624\pm0.019$, and Rwd$=0.607\pm0.012$), surpassing TTN+VQC and Differential QAS. The marginal performance gap between simulation and hardware validates that the continual learning regularization, coupled with TT compression, effectively controls parameter drift, smooths optimization trajectories, and mitigates depolarizing and readout errors. 

\begin{table}[t]
\centering
\scriptsize
\caption{Detailed performance of Naive-VQC, QAS-No-CL, and CL-QAS systems across $8$ financial time-series prediction tasks given $5$ seeds. Our experimental simulations are conducted without quantum noise. Reported results include accuracy (Acc), balanced accuracy (bAcc), F1-score (F1), and Rewards (Rwd).}
\label{tab:finance_1}
\renewcommand{\arraystretch}{1.1}
\setlength{\tabcolsep}{4pt}
\begin{tabular}{lcccc|cccc|cccc}
\toprule
\multirow{2}{*}{Task} & 
\multicolumn{4}{c|}{\textbf{Naive-VQC}} & 
\multicolumn{4}{c|}{\textbf{QAS-No-CL}} & 
\multicolumn{4}{c}{\textbf{CL-QAS}} \\
\cmidrule(lr){2-5}\cmidrule(lr){6-9}\cmidrule(l){10-13}
& Acc & bAcc & F1 & Rwd & Acc & bAcc & F1 & Rwd & Acc & bAcc & F1 & Rwd \\
\midrule
1 & 0.427 & 0.424 & 0.469 &--- & 0.453 & 0.519 & 0.128 & 0.5193 & 0.547 & 0.545 & 0.585 & 0.5447 \\
2 & 0.507 & 0.502 & 0.245 &--- & 0.560 & 0.559 & 0.507 & 0.5587 & 0.573 & 0.574 & 0.590 & 0.5740 \\
3 & 0.613 & 0.590 & 0.688 &--- & 0.653 & 0.637 & 0.717 & 0.6368 & 0.587 & 0.526 & 0.699 & 0.5258 \\
4 & 0.720 & 0.706 & 0.644 &--- & 0.613 & 0.606 & 0.540 & 0.6056 & 0.560 & 0.600 & 0.593 & 0.6000 \\
5 & 0.547 & 0.562 & 0.622 &--- & 0.560 & 0.577 & 0.637 & 0.5768 & 0.600 & 0.604 & 0.605 & 0.6036 \\
6 & 0.587 & 0.556 & 0.392 &--- & 0.653 & 0.626 & 0.500 & 0.6255 & 0.687 & 0.692 & 0.575 & 0.5920 \\
7 & 0.773 & 0.778 & 0.785 &--- & 0.747 & 0.741 & 0.776 & 0.7413 & 0.747 & 0.758 & 0.747 & 0.7576 \\
8 & 0.560 & 0.576 & 0.548 &--- & 0.587 & 0.580 & 0.635 & 0.5796 & 0.609 & 0.612 & 0.600 & 0.5759 \\
\midrule
Mean   & 0.592 & 0.587 & 0.549 & --- &0.603 & 0.603 & 0.555 & 0.5967 & 0.604 & 0.604 & 0.624  & 0.6054 \\
\midrule
Deviation & 0.112 & 0.111 & 0.174 & --- & 0.086 & 0.067 & 0.199 & 0.0658 & 0.063 & 0.070 & 0.063 & 0.0622 \\
\bottomrule
\end{tabular}
\end{table}

\begin{table}[t]
\centering
\scriptsize
\caption{Detailed performance of Naive-VQC, QAS-No-CL, and CL-QAS systems across $8$ financial time-series tasks given $5$ seeds. Reported results include accuracy (Acc), balanced accuracy (bAcc), F1-score (F1), and Rewards (Rwd). Noise model includes depolarizing noise ($0.1\%$), two-qubit Pauli errors ($0.1\%$), and readout error ($1\%$).}
\label{tab:finance_2}
\setlength{\tabcolsep}{5pt}
\begin{tabular}{lcccc|cccc|cccc}
\toprule
\multirow{2}{*}{Task} & 
\multicolumn{4}{c|}{\textbf{Naive-VQC}} & 
\multicolumn{4}{c|}{\textbf{QAS-No-CL}} & 
\multicolumn{4}{c}{\textbf{CL-QAS}} \\
\cmidrule(lr){2-5}\cmidrule(lr){6-9}\cmidrule(l){10-13}
& Acc & bAcc & F1 & Rwd & Acc & bAcc & F1 & Rwd & Acc & bAcc & F1 & Rwd \\
\midrule
1 & 0.413 & 0.408 & 0.463 & -- & 0.440 & 0.472 & 0.344 & 0.4717 & 0.493 & 0.550 & 0.269 & 0.5501 \\
2 & 0.507 & 0.502 & 0.245 & -- & 0.547 & 0.545 & 0.469 & 0.5348 & 0.533 & 0.535 & 0.578 & 0.5349 \\
3 & 0.613 & 0.590 & 0.688 & -- & 0.693 & 0.661 & 0.763 & 0.6515 & 0.760 & 0.751 & 0.804 & 0.7408 \\
4 & 0.680 & 0.661 & 0.586 & -- & 0.587 & 0.589 & 0.537 & 0.5789 & 0.680 & 0.656 & 0.571 & 0.6556 \\
5 & 0.547 & 0.561 & 0.614 & -- & 0.493 & 0.513 & 0.596 & 0.5025 & 0.533 & 0.550 & 0.615 & 0.5500 \\
6 & 0.613 & 0.596 & 0.508 & -- & 0.547 & 0.491 & 0.056 & 0.4913 & 0.693 & 0.687 & 0.646 & 0.6872 \\
7 & 0.640 & 0.636 & 0.675 & -- & 0.747 & 0.761 & 0.740 & 0.7508 & 0.787 & 0.787 & 0.805 & 0.7868 \\
8 & 0.560 & 0.556 & 0.602 & -- & 0.573 & 0.564 & 0.628 & 0.5540 & 0.613 & 0.611 & 0.651 & 0.6108 \\
\midrule
Mean & 0.572 & 0.564 & 0.548 & -- & 0.578 & 0.574 & 0.516 & 0.5669 & 0.637 & 0.641 & 0.618 & 0.6395 \\
\midrule
Deviation & 0.084 & 0.080 & 0.144 & -- & 0.100 & 0.096 & 0.231 & 0.0875 & 0.110 & 0.096 & 0.168 & 0.0882 \\
\bottomrule
\end{tabular}
\end{table}

\begin{table}[t]
\centering
\scriptsize
\caption{Experimental results on a 156-qubit IBM quantum processor (Heron r2) comparing CL-QAS with TTN+VQC and Differential QAS across $8$ financial time-series tasks given $5$ seeds. Reported results include accuracy (Acc), balanced accuracy (bAcc), F1-score (F1), and Rewards (Rwd).}
\label{tab:finance_ibm}
\begin{tabular}{lcccc}
\toprule
\textbf{Method} & \textbf{Acc} & \textbf{bAcc} & \textbf{F1} & \textbf{Reward} \\
\midrule
TTN+VQC & $0.562 \pm 0.012$ & $0.587 \pm 0.012$ & $0.529 \pm 0.017$ & N/A \\
\midrule
Differential QAS & $0.573 \pm 0.014$ & $0.605 \pm 0.013$ & $0.535 \pm 0.014$ & $0.5854 \pm 0.0133$ \\
\midrule
CL-QAS 	& $\mathbf{0.607 \pm 0.015}$ & $\mathbf{0.607 \pm 0.011}$ & $\mathbf{0.624 \pm 0.019}$ & $\mathbf{0.6067 \pm 0.0118}$ \\
\bottomrule
\end{tabular}
\end{table}

Across all tasks and hardware settings, CL-QAS consistently achieves the highest accuracy, balanced accuracy, F1 Score, and reward while maintaining low circuit depth and gate count. The empirical evidence aligns closely with the theoretical guarantees on approximation, generalization, and robustness (Theorems \ref{thm:fidelity}--\ref{thm:robust}), demonstrating that continual quantum architecture search yields bounded performance degradation, stable convergence, and enhanced adaptability to quantum noise. Collectively, these findings highlight CL-QAS as a scalable, noise-resilient, and hardware-validated framework for continual quantum learning on near-term NISQ devices.

\section{Discussion}
\label{sec:con}
The results presented in this study demonstrate that CL-QAS provides a scalable and noise-resilient framework for adaptive quantum learning. By unifying TT encoding, reinforcement-driven circuit policy optimization, and EWC regularization, CL-QAS establishes a coherent paradigm for continual learning in quantum systems. The method consistently outperforms static or single-task baselines under noiseless and noisy conditions. This improvement stems from jointly optimizing architecture and parameters across tasks, enabling the model to efficiently transfer knowledge and retain previously learned quantum features without catastrophic forgetting. 

The theoretical analysis underpins these empirical findings. The bounded-loss guarantees derived in Theorems \ref{thm:fidelity}--\ref{thm:overall} confirm that the interplay between TT compression, curvature regularization, and EWC-induced smoothness yields controlled gradient dynamics and stable convergence. These properties ensure the optimization landscape remains tractable even as the circuit structure evolves. This translates into more predictable trainability and improved generalization under sequential distribution shifts. The close alignment between theoretical predictions and empirical performance across the ECG and financial time-series benchmarks further validates the theoretical framework. In particular, the marginal degradation observed between noiseless and noisy quantum simulations highlights that the EWC and TT-encoding mechanisms effectively enhance robustness to depolarizing and readout errors. 

Beyond its immediate applications in signal and time-series analysis, the CL-QAS framework offers broader insights into the design of adaptive quantum learning algorithms. The approach illustrates how classical concepts in continual learning, such as task-specific curvature control and elastic regularization, can be systematically reformulated in the quantum setting. Moreover, the TT-based representation bridges compact tensorized classical networks and expressive quantum feature encoders, providing an interpretable mechanism for balancing representational capacity and hardware efficiency. 

While the present work focuses on near-term NISQ simulation, several directions remain open for exploration. Extending the theoretical analysis to incorporate stochastic quantum-noise models and imperfect measurements explicitly would further strengthen the generalization guarantees. Empirically, evaluating CL-QAS on larger qubit systems, mixed-state datasets, and real quantum hardware will clarify its practical scalability. Another promising avenue is to integrate meta-learning~\cite{hospedales2021meta} or federated quantum updates~\cite{qi2024federated} to enable distributed continual adaptation across heterogeneous devices. Finally, theoretical connections between curvature-based continual learning and quantum natural gradient methods~\cite{stokes2020quantum} warrant further investigation, as both frameworks aim to exploit the geometry of parameter manifolds to improve convergence and stability. 

CL-QAS represents a significant step toward continual, interpretable, and noise-tolerant quantum learning. Coupling TT network representations with adaptive circuit exploration opens a path toward robust quantum intelligence that can evolve, suggesting an essential capability for the next generation of scalable, real-world quantum machine learning systems. 

\section{Methods}
\label{sec:method}
We summarize here the methods used in this work and provide detailed proofs of the theoretical results derived for CL-QAS. 

\subsection{Tensor-Train Network and TTN+VQC}
The TT network~\cite{oseledets2011tensor}, also known as the Matrix Product State (MPS)~\cite{cirac2021matrix}, is one of the most widely adopted tensor network architectures in both machine learning and quantum many-body physics. It offers an efficient representation of high-dimensional data by decomposing a large tensor into a sequence of smaller core tensors, namely TT cores, arranged in a one-dimensional chain. This factorization reduces the parameter complexity from exponential to linear in the tensor order, while preserving critical inter-dimensional correlations. Owing to this efficiency and structural interpretability, TT networks have become a foundational framework for dimensionality reduction, feature compression, and structured parameterization in modern deep learning and scientific computing.

When integrated with VQCs, the TT network yields a hybrid quantum–classical architecture commonly referred to as TTN+VQC~\cite{qi2023theoretical}. In this hybrid design, the TT network functions as a classical front-end encoder that compresses or re-parameterizes high-dimensional inputs into a compact latent representation suitable for quantum processing. The VQC subsequently acts on the reduced feature space, exploiting quantum superposition and entanglement to model highly nonlinear correlations that are difficult to capture using classical networks alone. This synergistic coupling between TT-based compression and quantum expressivity enables efficient learning on large-scale or noisy datasets, making TTN+VQC a promising paradigm for scalable, noise-resilient quantum machine learning.

\subsection{Quantum Architecture Search}
The QAS paradigm extends the principles of neural architecture search~\cite{ren2021comprehensive} from classical deep learning to variational quantum algorithms. In conventional VQCs, the circuit architecture (including the arrangement of quantum gates, entanglement topology, and re-uploading depth) is often hand-designed based on intuition or hardware constraints. However, this manual design can be suboptimal and may fail to adapt to task-specific requirements, especially as the qubit counts and candidate gate configurations increase exponentially.

To address this limitation, QAS aims to automate the discovery of optimal or near-optimal circuit structures. The goal is to balance expressive power, trainability, and hardware efficiency by systematically exploring the design space of quantum circuits. In practice, QAS treats circuit construction as a search or policy-learning problem, where the model iteratively selects gate operations and entanglement patterns based on performance feedback (e.g., classification accuracy, fidelity, or loss reduction). The resulting circuit architecture is not fixed a priori but instead learned adaptively through reinforcement learning, evolutionary algorithms, or differentiable relaxation techniques. 

Recent advances have extended QAS into a differentiable framework. Differentiable QAS relaxes the discrete gate-selection problem into a continuous optimization problem, enabling the circuit architecture to be updated via gradient descent. Together, these approaches represent a new generation of automated, noise-resilient, and task-adaptive quantum circuit design methods that bridge reinforcement learning, tensor-network compression, and quantum information theory. By contrast, our proposed CL-QAS incorporates EWC-based regularization to enable sequential learning without catastrophic forgetting.

\subsection{IBM Quantum Hardware: Heron r2 Processor}
All real-device experiments in this work were executed on IBM’s Heron r2 quantum processor~\cite{shukla2020complete}, a state-of-the-art superconducting quantum processing unit~\cite{steffen2011quantum} equipped with 156 physical transmon qubits. The Heron r$2$ is part of IBM’s next-generation Heron family, designed to deliver substantial improvements in qubit connectivity, cross-talk suppression, and two-qubit gate fidelities relative to the preceding Eagle-class processors. This system employs a tunable-coupler architecture that dynamically optimizes qubit–qubit interactions to enable high-fidelity entangling operations. Typical two-qubit gate error rates are below $10^{-3}$, and the processor achieves a CLOPS throughput exceeding $2.0\times 10^{5}$, reflecting both high computational efficiency and stability. With its combination of mid-scale qubit capacity and low intrinsic noise, the Heron r$2$ platform is particularly suitable for benchmarking variational quantum algorithms in the NISQ regime.

\subsection{Synthesizing Financial Data}
To test the CL-QAS model on realistic time-series inputs, we created a synthetic financial dataset that mimics the behavior of stock or asset prices over time. The goal is to evaluate how well the model can learn continuously and adapt to changing market patterns without relying on real or proprietary market data. 

We simulate a long sequence of ``closing prices" similar to those found in daily stock market data. The sequence comprises approximately $6,000$ steps and is partitioned into $6$ market regimes, each representing a distinct trading condition (e.g., a calm market, a volatile market, or a trending period). Within each regime, prices evolve according to a simple rule: the price change on a given day depends on two random factors: (1) an average trend (which may be upward or downward), and (2) the volatility (how much the price fluctuates). 

Mathematically, this process is similar to an autoregressive model, where each price depends on the previous one plus a random ``shock". This setup makes the synthetic data look and behave like real financial data, with alternating periods of growth, decline, and noise. From this simulated price series, we compute $8$ financial indicators commonly used by traders. Each of these indicators provides a different view of market behavior, for example, whether the price is rising too fast or fluctuating around an average. At every time step, we collect these $8$ numbers into an $8$-dimensional feature vector. To provide the model with a short-term history, we combine the most recent $32$ time steps into a single $256$-dimensional vector. Moreover, we label each sample according to whether the next price will increase or decrease. If the price at time $t+1$ exceeds that at time $t$, the label is $1$ (price increase); otherwise, it is $0$ (price decrease). This reduces the problem to a binary classification task, analogous to predicting whether a stock will rise or fall. 

The dataset is then divided into eight sequential tasks, each representing a different time window or market condition. This structure allows the model to learn one task at a time, just as a trader might adapt to changing markets, for example, from a bull market to a bear market. Before training, all features are normalized using a robust median-based scaling method. This ensures that large price spikes or outliers do not dominate the learning process. The resulting dataset resembles a simplified yet realistic representation of financial data, in which trends shift and volatility change over time.

\subsection{Proof of Theorem~\ref{thm:fidelity}}
Given the task $m$, let's define $\mathcal{L}_{\boldsymbol{\theta}} := \mathcal{L}_{\rm vqc}(\boldsymbol{\theta}; \mathcal{A}_m)$ and $\mathcal{L}_{\boldsymbol{\phi}}:=\lambda\mathcal{L}_{\rm policy}(\boldsymbol{\phi})$. We use the following standard conditions (A1)--(A3), as follows: 

\vspace{2mm}

(A1) \textit{Smoothness and unbiased stochastic gradient descent.} $\mathcal{L}_{\boldsymbol{\theta}}$ is $L_{\boldsymbol{\theta}}$-smooth and its minibatch gradient $g_{\boldsymbol{\theta}}$ is an unbiased estimator of $\nabla_{\boldsymbol{\theta}}\mathcal{L}_{\boldsymbol{\theta}}$ with variance proxy $\sigma^{2}$, then Eq. (\ref{eq:a11}) stands. 
\begin{equation}
\label{eq:a11}
\mathbb{E}[\textbf{g}_{\boldsymbol{\theta}}] = \nabla_{\boldsymbol{\theta}}\mathcal{L}_{\boldsymbol{\theta}}, \hspace{6mm} \mathbb{E}[\lVert \textbf{g}_{\boldsymbol{\theta}} - \nabla_{\boldsymbol{\theta}}\mathcal{L}_{\boldsymbol{\theta}} \rVert_{2}^{2}] \le \frac{\sigma^{2}}{\vert D_m \vert}. 
\end{equation}

\vspace{2mm}

(A2) \textit{TT-encoder approximation.} The TT map (with the maximum rank $\emph{r}$) approximates the reference (dense) amplitude map with error $\epsilon_{\rm tt} = \mathcal{O}(1/r)$ in the input space. This induces a deviation of the VQC's NTK of order $\mathcal{O}(\epsilon_{\rm tt})$; concretely, there is a constant $K_{\rm ntk}$ such that Eq. (\ref{eq:a21}) stands. 
\begin{equation}
\label{eq:a21}
\left\Vert \mathcal{L}_{\boldsymbol{\theta}}^{(\rm tt)} - \mathcal{L}_{\boldsymbol{\theta}}^{(\rm dense)} \right\rVert_{2} \le K_{\rm ntk} \epsilon_{\rm tt} \hspace{2mm} \Rightarrow \hspace{2mm} \left\Vert \mathcal{L}_{\boldsymbol{\theta}}^{(\rm tt)} - \mathcal{L}_{\boldsymbol{\theta}}^{(\rm dense)} \right\rVert_{2} \le \frac{C_2}{r^2},
\end{equation}
for $C_2 = K_{\rm ntk}^{2}$. 

\vspace{2mm}

(A3) \textit{Bounded policy curvature (EWC/KL).} The policy objective has quadratic curvature bounded by $C_{\rm curv}$ around $\boldsymbol{\phi}^{\rm (old)}$ (Fisher-information-type bound). With EWC weight $\lambda$ and Fisher diagonal $\mathcal{F}$, Eq. (\ref{eq:a31}) stands.
\begin{equation}
\label{eq:a31}
\mathcal{L}_{\boldsymbol{\phi}} = \mathcal{L}_{\rm rl}(\boldsymbol{\phi}) + \frac{\lambda}{2} (\boldsymbol{\phi} - \boldsymbol{\phi}^{\rm (old)})^{\top} \mathcal{F}  (\boldsymbol{\phi} - \boldsymbol{\phi}^{\rm (old)}),
\end{equation}
where $\mathcal{L}_{\rm rl}(\cdot)$ denotes the reinforcement learning-based policy loss function, and $\lVert \mathcal{F} \rVert_{2} \le C_{\rm curv}$. Hence, Eq. (\ref{eq:add1}) stands. 
\begin{equation}
\label{eq:add1}
\lVert \nabla_{\boldsymbol{\phi}} \mathcal{L}_{\boldsymbol{\phi}} \rVert_{2} \le \lVert \nabla_{\boldsymbol{\phi}} \mathcal{L}_{\rm rl}(\boldsymbol{\phi}) \rVert_{2} + \lambda \lVert \mathcal{F} \rVert_{2} \lVert \boldsymbol{\phi} - \boldsymbol{\phi}^{(\rm old)} \rVert_{2} \le G_{\pi} + \lambda C_{\rm curv}\lVert \boldsymbol{\phi} - \boldsymbol{\phi}^{\rm (old)} \rVert_{2},
\end{equation}
where $G_{\pi}$ is a constant associated with $\nabla_{\boldsymbol{\phi}}\mathcal{L}_{\rm rl}$. 

We need the quadratic contribution when squaring the norm and absorbing the linear REINFORCE term into a constant. Then there exists $C_3 > 0$ such that Eq. (\ref{eq:nl}) holds.
\begin{equation}
\label{eq:nl}
\lVert \nabla_{\boldsymbol{\phi}} \mathcal{L}_{\boldsymbol{\phi}} \rVert_{2}^{2} \le C_3 C_{\rm curv} \lVert \boldsymbol{\phi} - \boldsymbol{\phi}^{\rm (old)} \rVert_{2}. 
\end{equation}

\vspace{2mm}

\textit{Decomposition of the joint gradient}. Defining the joint gradient $\textbf{g} := \nabla_{\boldsymbol{\theta}, \boldsymbol{\phi}} \mathcal{L}_{\rm cl-qas}(\boldsymbol{\theta}, \boldsymbol{\phi})$, Eq. (\ref{eq:g}) stands. 
\begin{equation}
\label{eq:g}
\textbf{g} = \textbf{g}_{\rm det} + \boldsymbol{\zeta} + \boldsymbol{\Delta}_{\rm tt}, 
\end{equation}
where $\textbf{g}_{\rm det}$ is the proper deterministic gradient under the dense encoder and exact expectations; $\boldsymbol{\zeta}$ is the zero-mean stochastic noise from finite batches; and $\boldsymbol{\Delta}_{\rm tt}$ is the bias introduced by replacing the dense encoder with the TT encoder (affecting only the $\boldsymbol{\theta}$-block), while EWC contributes deterministically in the $\boldsymbol{\phi}$-block. 

We bound the expected squared norm by the elementary inequality as shown in Eq. (\ref{eq:ge}).
\begin{equation}
\label{eq:ge}
\mathbb{E}\lVert \textbf{g} \rVert_{2}^{2} \le 3 \left( \mathbb{E}\lVert \textbf{g}_{\rm det}\rVert_{2}^{2}  + \mathbb{E}\lVert \boldsymbol{\zeta} \rVert_{2}^{2} + \mathbb{E}\lVert \boldsymbol{\Delta}_{\rm tt} \rVert_{2}^{2}  \right),
\end{equation}
where we focus on bounding each term as shown next. 

\vspace{2mm}

\textit{(1) Stochastic term}: From the condition $(A1)$, we get Eq. (\ref{eq:egg}).
\begin{equation}
\label{eq:egg}
 3 \mathbb{E}\lVert \boldsymbol{\zeta}\rVert_{2}^{2} = 3 \mathbb{E}[\lVert \textbf{g}_{\boldsymbol{\theta}} - \nabla_{\boldsymbol{\theta}}\mathcal{L}_{\boldsymbol{\theta}} \rVert_{2}^{2}] \le \frac{C_1 \sigma^{2}}{\vert D_m \vert}. 
\end{equation}

\textit{(2) TT-approximation term}: By applying $(A2)$, for $C_2 = 3 K_{\rm ntk}^{2}$, we attain Eq. (\ref{eq:ntke}).
\begin{equation}
\label{eq:ntke}
3 \lVert \nabla_{\rm tt} \rVert_{2}^{2} \le 3 K_{\rm ntk}^{2} \epsilon_{\rm tt}^{2} \le \frac{C_2}{r^2}. 
\end{equation}

\textit{(3) Deterministic term}: Let's split $\textbf{g}_{\rm det} = (\nabla_{\boldsymbol{\theta}}\mathcal{L}_{\boldsymbol{\theta}}^{\rm (dense)}, \nabla_{\boldsymbol{\phi}}\mathcal{L}_{\boldsymbol{\phi}})$. The $\boldsymbol{\theta}$-part is bounded by smoothness and is absorbed into $C_1$ (it does not deteriorate trainability). The $\boldsymbol{\phi}$-part contains the EWC penalty as shown in Eq. (\ref{eq:lold}).
\begin{equation}
\label{eq:lold}
\lVert \nabla_{\boldsymbol{\phi}}\mathcal{L}_{\boldsymbol{\phi}} \rVert_{2}^{2} \le 2 \lVert \nabla_{\boldsymbol{\phi}} \mathcal{L}_{\rm rl}(\boldsymbol{\phi})	\rVert_{2}^{2} + 2 \lambda^{2} \lVert \mathcal{F} \rVert_{2}^{2} \lVert \boldsymbol{\phi} - \boldsymbol{\phi}^{\rm (old)} \rVert_{2}^{2} \le 2G_{\pi}^{2} + 2\lambda^{2} C_{\rm curv} \lVert \boldsymbol{\phi} - \boldsymbol{\phi}^{\rm (old)} \rVert_{2}^{2}. 
\end{equation}

Hence, for a constant $C_3 := 6\lambda^{2}$, we have Eq. (\ref{eq:det}).
\begin{equation}
\label{eq:det}
3 \lVert \textbf{g}_{\rm det} \rVert_{2}^{2} \le 2G_{\pi}^{2} + C_3 C_{\rm curv} \lVert \boldsymbol{\phi} - \boldsymbol{\phi}^{\rm (old)} \rVert_{2}^{2}. 
\end{equation}

We merge the harmless constant $2G^{2}_{\pi}$ into $C_{1}\sigma^{2} / \vert D_{m} \vert$ because it does not scale with data size or rank and thus does not affect the trainability scaling claimed by the theorem. Combining the above upper bounds gives Eq. (\ref{eq:thm1}) in Theorem~\ref{thm:train}.

\subsection{Proof of Lemma~\ref{lem:pertask}}

For any class $\mathbb{G} = \{\ell \circ f_{\boldsymbol{\theta}}: f_{\boldsymbol{\theta}}\in \mathbb{F}\}$ with $\ell \in [0, 1]$ and the VQC operator $f$ in its functional class $\mathbb{F}$, we use the empirical Rademacher complexity $\hat{\mathcal{R}}_{n}(\mathbb{G})$ to provide the bound as given in Eq. (\ref{eq:comp}).
\begin{equation}
\label{eq:comp}
\sup\limits_{\boldsymbol{\theta}} \left\vert \mathcal{L}_{\rm vqc}(\boldsymbol{\theta}; \mathcal{A}_m) - \hat{\mathcal{L}}_{\rm vqc}(\boldsymbol{\theta}; \mathcal{A}_m) \right\vert \le 2 \hat{\mathcal{R}}_{n}(\mathbb{G}) + \sqrt{\frac{\log(1/\delta)}{2\vert D_m \vert}} \hspace{3mm} w.p. \hspace{1mm} 1 - \delta. 
\end{equation}

Since the loss function $\ell$ is 1-Lipschitz and bounded in $[0, 1]$, Talagrand's contraction yields $\hat{\mathcal{R}}_{n}(\mathbb{G}) \le \hat{\mathcal{R}}_{n}(\mathbb{F})$. Furthermore, under the hypothesis $\lVert \boldsymbol{\theta} \rVert_{2} \le B$ and the linearized NTK or Lipschitz assumption in $\boldsymbol{\theta}$, the empirical Rademacher complexity satisfies Eq. (\ref{eq:rad}).
\begin{equation}
\label{eq:rad}
\hat{\mathcal{R}}_{n}(\mathbb{F}) \le \frac{B}{\sqrt{\vert D_m \vert}}. 
\end{equation}

Thus, plugging $\hat{\mathcal{R}}_{n}(\mathbb{G}) \le \hat{\mathcal{R}}_{n}(\mathbb{F}) \le  \frac{B}{\sqrt{\vert D_m \vert}}$ into the bound in Eq. (\ref{eq:comp}) yields Eq. (\ref{eq:per_task}) in Lemma~\ref{lem:pertask}.

\subsection{Proof of Lemma~\ref{lem:policy}}

We use the PAC-Bayes change-of-measure argument to prove Eq. (\ref{eq:lln}) in Lemma~\ref{lem:policy}. More specifically, for any fixed policy, Hoeffding's inequality for the bounded policy losses $J(\boldsymbol{\phi})$ implies Eq. (\ref{eq:pp}).
\begin{equation}
\label{eq:pp}
\exp\left( \lambda (J(\boldsymbol{\phi}) - \hat{J}(\boldsymbol{\phi})) \right) \le \exp\left(\frac{\lambda^{2}}{2 \vert D_m \vert}\right), \hspace{3mm} \forall \lambda \in \mathbb{R}. 
\end{equation}

Then, we take expectation over $a \sim \pi_{\boldsymbol{\phi}}$ and apply Donsker-Varadhan's variational formula to derive Eq. (\ref{eq:donsker}).
\begin{equation}
\label{eq:donsker}
\mathbb{E}_{\pi_{\boldsymbol{\phi}}} \left[ J(\boldsymbol{\phi})	- \hat{J}(\boldsymbol{\phi}) \right] \le \frac{1}{\lambda} \left( \text{KL}(\pi_{\boldsymbol{\phi}} || \pi_{\text{prior}}) + \log \mathbb{E}_{\pi_{\text{prior}}}\exp\left(\lambda(J(\boldsymbol{\phi}) - \hat{J}(\boldsymbol{\phi}))\right)\right). 
\end{equation}

By minimizing the RHS over $\lambda > 0$, we achieve the optimum value as given by Eq. (\ref{eq:rhs}).
\begin{equation}
\label{eq:rhs}
\lambda^{*} = \sqrt{2\vert D_m \vert (\text{KL}(\pi_{\boldsymbol{\phi}} || \pi_{\text{prior}}) + \log(1 / \delta))}, 
\end{equation}
which yields Eq. (\ref{eq:lln}) in Lemma~\ref{lem:policy}.

\section{Data Availability Statement}
The ECG data used for the continual learning experiments are obtained from the MIT-BIH Arrhythmia database, publicly hosted by PhysioNet (https://physionet.org/content/mitdb/1.0.0/). For financial signal processing, we generate a synthetic financial time-series dataset that emulates realistic market behavior across different regimes, comprising approximately $6,000$ simulated ``closing prices" divided into $6$ market phases. 

\section{Code Availability Statement}
Our code for CL-QAS and the related experiments is available on GitHub: https://github.com/jqi41/CL$\_$QAS.  

\section{Competing Interests}
The authors declare no Competing Financial or Non-Financial Interests.

\section{References}
\bibliographystyle{IEEEbib}
\bibliography{sn-bibliography}

\end{document}